\documentclass[
 aip,
 long, numerical bibliography, (default)
 jmp,%
 amsmath,amssymb,
reprint,%
superscriptaddress,
nofootinbib
]{revtex4-1}
\usepackage{graphicx}% Include figure files
\usepackage{dcolumn}% Align table columns on decimal point
\usepackage{bm}% bold math
\usepackage{epstopdf}
\usepackage{xr}
\usepackage{amsmath}
\usepackage{color}
\usepackage[normalem]{ulem}
\usepackage{wasysym}
\usepackage{siunitx}
\usepackage{physics}
\usepackage{lineno}

\usepackage[T1]{fontenc}
\usepackage{booktabs}
\usepackage[section]{placeins}

\usepackage{textcomp}
\makeatletter

\newcommand{\Rmnum}[1]{\expandafter\@slowromancap\romannumeral #1@}
\makeatother

\definecolor{mygreen1}{rgb}{0, 0.5, 0.7}

\begin{document}
\markboth{\jobname}{\jobname .tex}
%\preprint{AIP/123-QED}

\title{Non-trivial quantum oscillation geometric phase shift in a trivial band}
\author{Biswajit Datta}
\affiliation{Department of Condensed Matter Physics and Materials Science, Tata Institute of Fundamental Research, Homi Bhabha Road, Mumbai 400005, India}
\homepage{biswajit.datta@tifr.res.in}
\author{Pratap Chandra Adak}
\affiliation{Department of Condensed Matter Physics and Materials Science, Tata Institute of Fundamental Research, Homi Bhabha Road, Mumbai 400005, India}
\author{Li-kun Shi}
\affiliation{Institute of High Performance Computing, Agency for Science, Technology, \& Research, Singapore 138632}
\author{Kenji Watanabe}
\affiliation{National Institute for Materials Science, 1-1 Namiki, Tsukuba 305-0044, Japan}
\author{Takashi Taniguchi}
\affiliation{National Institute for Materials Science, 1-1 Namiki, Tsukuba 305-0044, Japan}
\author{Justin C. W. Song}
\affiliation{Institute of High Performance Computing, Agency for Science, Technology, \& Research, Singapore 138632}
\affiliation{Division of Physics and Applied Physics, Nanyang Technological University, Singapore 637371}
\author{Mandar M. Deshmukh}
\homepage{deshmukh@tifr.res.in}
\affiliation{Department of Condensed Matter Physics and Materials Science, Tata Institute of Fundamental Research, Homi Bhabha Road, Mumbai 400005, India}
%\date{\today}

%abstract

\begin{abstract}

The accumulation of non-trivial geometric phases in a material's response is often a tell-tale sign of a rich underlying internal structure~\cite{RevModPhys.82.1959,analytis2010two,mikitik_manifestation_1999}. Studying quantum oscillations provides one of the ways to determine these geometrical phases, such as Berry's phase~\cite{pancharatnam1956s,berry1984quantal,PhysRevLett.93.166402}, that play a central role in topological quantum materials. We report on magneto-transport measurements in ABA-trilayer graphene, the band structure of which is comprised of
a weakly gapped linear Dirac band, nested within a trivial quadratic
band~\cite{taychatanapat_quantum_2011,serbyn_new_2013,datta_strong_2017}. Here we show Shubnikov-de Haas
(SdH) oscillations of the quadratic band shifted by a phase that sharply departs from the expected 2$\pi$
Berry's phase. Our analysis reveals that, surprisingly, the anomalous phase shift is non-trivial and is inherited
from the non-trivial Berry's phase of the linear Dirac band due to strong filling-enforced constraints between
the linear and quadratic band Fermi surfaces. Given that many topological materials contain multiple bands,
our work indicates how additional bands, which are thought to obscure the analysis, can actually be exploited
to tease out the subtle effects of Berry's phase.

\end{abstract}

\maketitle

Non-trivial geometric phases can arise from diverse settings including in strong spin-orbit coupled systems that possess real-space~\cite{muhlbauer2009skyrmion,nagaosa2013topological} or momentum-space spin-texture~\cite{bernevig2013topological}, periodic driving by strong electromagnetic fields~\cite{lindner2011floquet}, and multi-orbital/site structure within a unit cell~\cite{xiao2007valley}. Even though such phases are often encoded in the subtle twisting of electronic wavefunctions, their impact on material response can be profound, being responsible for a wealth of unconventional quantum behaviors that include unconventional magneto-electric coupling~\cite{lee2017valley}, an emergent electro-magnetic field for electrons~\cite{nagaosa2013topological}, and protected edge modes~\cite{ryu2002topological} amongst others.

A prominent example is the Berry's phase~\cite{pancharatnam1956s,berry1984quantal,PhysRevLett.93.166402}. In anomalous Hall metals, the Berry's phase on the Fermi surface determines the (un-quantized part of the) anomalous Hall conductivity~\cite{haldane2004berry,RevModPhys.82.1539}; non-trivial $\pi$ Berry's phase enforces the absence of back-scattering in topological materials~\cite{ando1998berry}. Indeed, the value of the Berry's phase of electrons as they encircle a single, closed Fermi surface can be used as a litmus-test for topological bands ---  $\pi$ indicates a non-trivial band~\cite{novoselov_two-dimensional_2005,zhang_experimental_2005,koshino_trigonal_2009, zhang_experimental_2011,buttner2011single}, whereas 2$\pi$ indicates a trivial band~\cite{novoselov_unconventional_2006, Marzari_2011,mikitik_electron_2008}. In the presence of a magnetic field ($B$), the (quantized) size of closed cyclotron orbits depends on both the magnetic flux threading the orbits as well as the Berry's phase of electrons. As a result, quantum oscillations of a closed Fermi surface can acquire phase shifts -- a direct result of the Berry's phase of electrons~\cite{mikitik_manifestation_1999}. This is visible in oscillations of both resistance and thermodynamic quantities like magnetization. Tracking such quantum oscillations phase shifts have emerged as a powerful probe for topological materials~\cite{murakawa2013detection,wang_topological_2015,akiyama_berry_2017,ghahari_off_2017,wang_topological_2015,rode_berry_2016}.

Here we unveil a new phase shift for quantum oscillations that appears in multi-Fermi-surface metals. In particular, we reveal how the quantum oscillations of a completely trivial Fermi surface (with a constant and trivial Berry's phase) can acquire non-trivial ($\pm\pi$) phase shifts that are gate-tunable. The anomalous phase shifts are found in measured SdH oscillations of a trivial band in a multi-band system -- ABA-trilayer graphene -- and, as we discuss below originate from strong filling-enforced-constraints among Fermi surfaces that unavoidably arise in multi-Fermi-surface metals. Our experiment probes for the first time the continuous variation of the Berry's phase induced quantum oscillation phase shift, as a function of gate voltage ($V_\mathrm{BG}$), in an inversion symmetry broken system close to the band edge.

%\textbf{\textcolor{mygreen1}{Fig.1a}}
%\textbf{4. Explain the specific device/system you made}

\begin{figure}
\includegraphics[width=8cm]{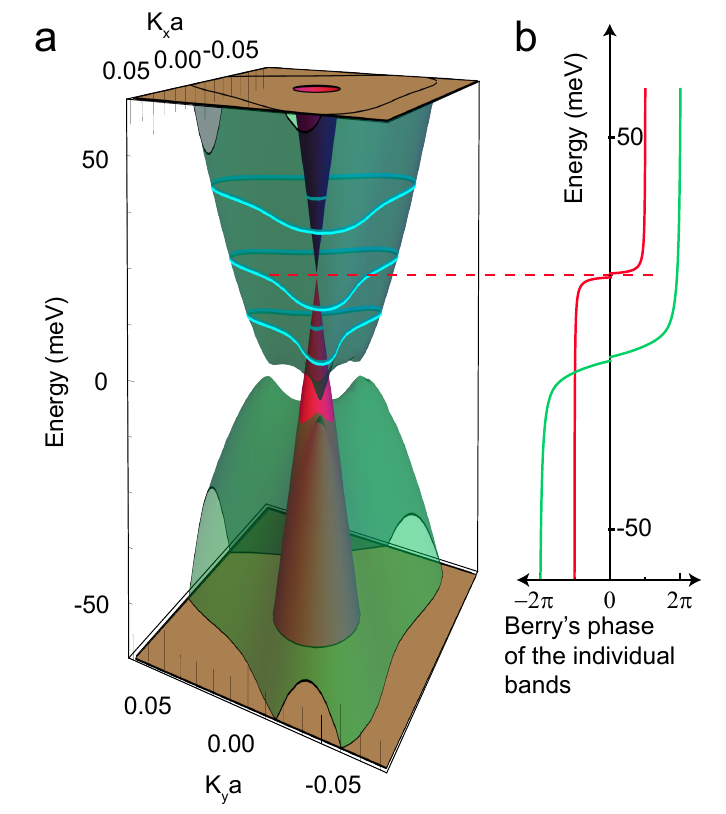}
\caption{ \label{fig:fig1} {\footnotesize \textbf{Band diagram of ABA-trilayer graphene.} (a) Band diagram of ABA-stacked trilayer graphene showing one pair of the conical band (colored red) and another pair of the quadratic band (colored green). Fermi surface at three different energies is overlaid for which Fermi energy lies in the valence band, band gap and in the conduction band of the MLG-like band. There is no contour from the MLG-like band when Fermi energy is in the band gap. (b) Calculated Berry's phase plot with same color codes for both the bands. Since the bands are gapped, Berry's phase of the individual bands goes to zero at the respective band edges.}}
\end{figure}

We study a high mobility hexagonal boron nitride (hBN) encapsulated ABA-stacked trilayer graphene device (see Supplementary Materials~\Rmnum{1}). A metal top gate and a highly doped silicon back gate ensure independent tunability of charge carrier density and electric field. All the measurements are done with a low-frequency lock-in technique at 1.5~K. ABA-trilayer graphene is very interesting because it is the simplest system supporting the simultaneous existence of a monolayer graphene (MLG)-like linear and a bilayer graphene (BLG)-like quadratic
band in experimentally accessible Fermi energy (see Fig.~\ref{fig:fig1}a)~\cite{koshino_landau_2011,serbyn_new_2013}. Broken inversion symmetry in ABA-trilayer graphene generates a small mass term in the Hamiltonian~\cite{koshino_landau_2011,serbyn_new_2013}. As a result, both the pairs of bands are individually gapped as seen in Fig.~\ref{fig:fig1}a; the band gap of the MLG-like Dirac cone~\cite{datta_landau_2018} is $\sim$1~meV. Fig.~\ref{fig:fig1}a shows that when both these bands are filled, the Fermi surface of the ABA-trilayer graphene consists of two Fermi contours-- the inner contour (smaller in size) comes from the MLG-like band and the outer contour (larger in size) comes from the BLG-like band. Fig.~\ref{fig:fig1}b shows that the MLG-like Dirac cone has a robust $\pi$ Berry's phase which only reduces to zero in the vicinity of the MLG-like band edge. On the other hand, the BLG-like conduction band has more-or-less a constant trivial Berry's phase $2\pi$ in the region of interest (around the Dirac cone gap).
In our experiment, we probe a narrow energy window near the MLG-like band gap. In the following, we use
``band gap'' to refer to the MLG-like Dirac cone gap.

\begin{figure*}
\includegraphics[width=15.5cm]{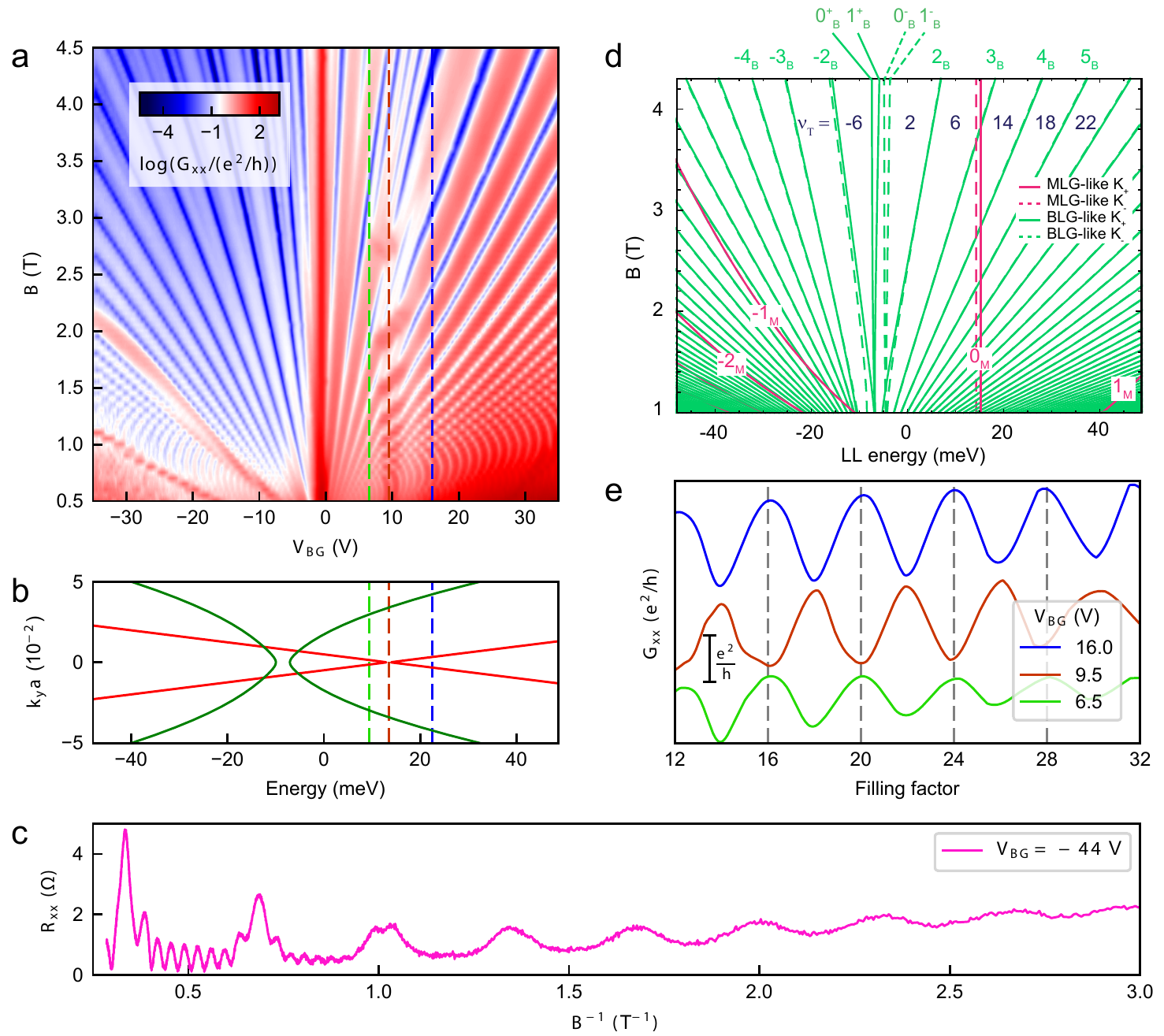}
\caption{ \label{fig:fig2} {\footnotesize \textbf{Magnetotransport of ABA-trilayer graphene.} (a) Colour scale plot of $G_{\mathrm{xx}}$ as a function of back gate voltage and magnetic field. The vertical feature parallel to the magnetic field axis at V$_{\mathrm{BG}}\sim10~V$ corresponds to the LL crossings of N$_\mathrm{M}$=0 LL with other BLG-like LLs. This V$_{\mathrm{BG}}$ also corresponds to the band gap of the MLG-like bands. (b) Energy band diagram shown in the same energy range as the experimental fan diagram shown in panel (a). (c) An SdH oscillation line slice showing the beating pattern due to the two bands with different Fermi surface areas. Low and high-frequency oscillations come from the MLG-like and the BLG-like bands respectively. (d) Theoretically calculated LL energies of the spin degenerate Landau levels as a function of magnetic field. Red and green lines denote LLs originating from the Dirac and the quadratic bands respectively. Solid and dashed lines denote LLs from K$_+$ and K$_-$ valleys respectively. (e) SdH oscillations (G$_\mathrm{xx}$) as a function of filling factor below the band gap (green), in the band gap (red) and above the band gap (blue) which show that the phase of the SdH oscillation in the band gap is $\pi$ shifted compared to the other two. The curves are shifted in the vertical direction for clarity. Gate voltage and approximate energy locations of the three SdH oscillation slices are marked with dashed lines of corresponding color in the fan diagram (a) and in the bandstructure (b) respectively.}}
\end{figure*}

%\textbf{\textcolor{mygreen1}{Fig.2}}
%\textbf{5. What happens at finite magnetic field}

In the presence of a magnetic field, the continuous band structure shown in Fig.~\ref{fig:fig1}a splits into Landau levels (LLs). The closed orbits $\vec{k}$ space area takes on quantized values that depend on the Berry's phase (and magnetic flux). As the magnetic field is swept and the charge density is varied independently, LLs cross the Fermi surface giving rise to the density of states oscillations that result in longitudinal conductance ($G_\mathrm{xx}$) oscillations~\cite{Smrcka_1986}. At a fixed density, the conductance oscillations (SdH) can be written as $\Delta G_{\mathrm{xx}}=G \cos [2\pi (\frac{B_\mathrm{F}}{B}+\gamma)]$ where G is the oscillation magnitude, $B_\mathrm{F}=\frac{n_\mathrm{S}h}{g e}$ is the SdH oscillation frequency in $1/B$ parameter space and the phase shift $\gamma=\frac{\mathrm{\Phi_\mathrm{B}}}{2\pi}-\frac{1}{2}$. Here, n$_\mathrm{S}$ is the density in S sub-band for a multiband system, $g$ is the LL degeneracy which is 4 for graphene, and $\Phi_\mathrm{B}$ is the Berry's phase. Fig.~\ref{fig:fig2}a shows~\cite{biswajit_datta_2018_1451851} our measured SdH oscillation in $G_{\mathrm{xx}}$ as a function of $B$ and $V_\mathrm{BG}$.  The corresponding band structure at zero magnetic field is shown in Fig.~\ref{fig:fig2}b. Fig.~\ref{fig:fig2}c shows an SdH oscillation with two distinct frequencies which reveal that two distinct Fermi surfaces are involved in the transport. Fig.~\ref{fig:fig2}a is composed of many such SdH oscillation slices at different gate voltages. Theoretically calculated LL diagram (Fig.~\ref{fig:fig2}d) shows that the MLG-like and the BLG-like LLs disperse as $\sim \sqrt{B}$ and $\sim B$ respectively~\cite{koshino_landau_2011,serbyn_new_2013,datta_strong_2017}. This distinct dispersion of the LLs along with the corresponding Hall conductance enable easy identification of the MLG-like and the BLG-like LLs~\cite{datta_strong_2017,datta_landau_2018,stepanov_tunable_2016,campos_landau_2016}.

%\textbf{6. Show anomalous phase}

The central result of our study -- that of an anomalous phase shift in the trivial BLG-like band -- is vividly illustrated in Fig.~\ref{fig:fig2}e. It shows three slices of BLG-like SdH oscillations at different densities away from the crossing points which correspond to Fermi levels in the valence band, in the gap, and in the conduction band of the MLG-like Dirac cone respectively. We emphasize that for all these three densities, the Fermi levels lie in the conduction band of the BLG-like band. The SdH oscillations above and below the gap clearly show a $\pi$ phase shift from the SdH oscillation at the gap. This is surprising since the BLG-like band in this energy range has a constant trivial Berry's phase (see Fig.~\ref{fig:fig1}b).

%\textbf{\textcolor{mygreen1}{LL fit details and extraction of Berry's phase}}

%LL fit details and extraction of Berry's phase

\begin{figure*}
\includegraphics[width=16cm]{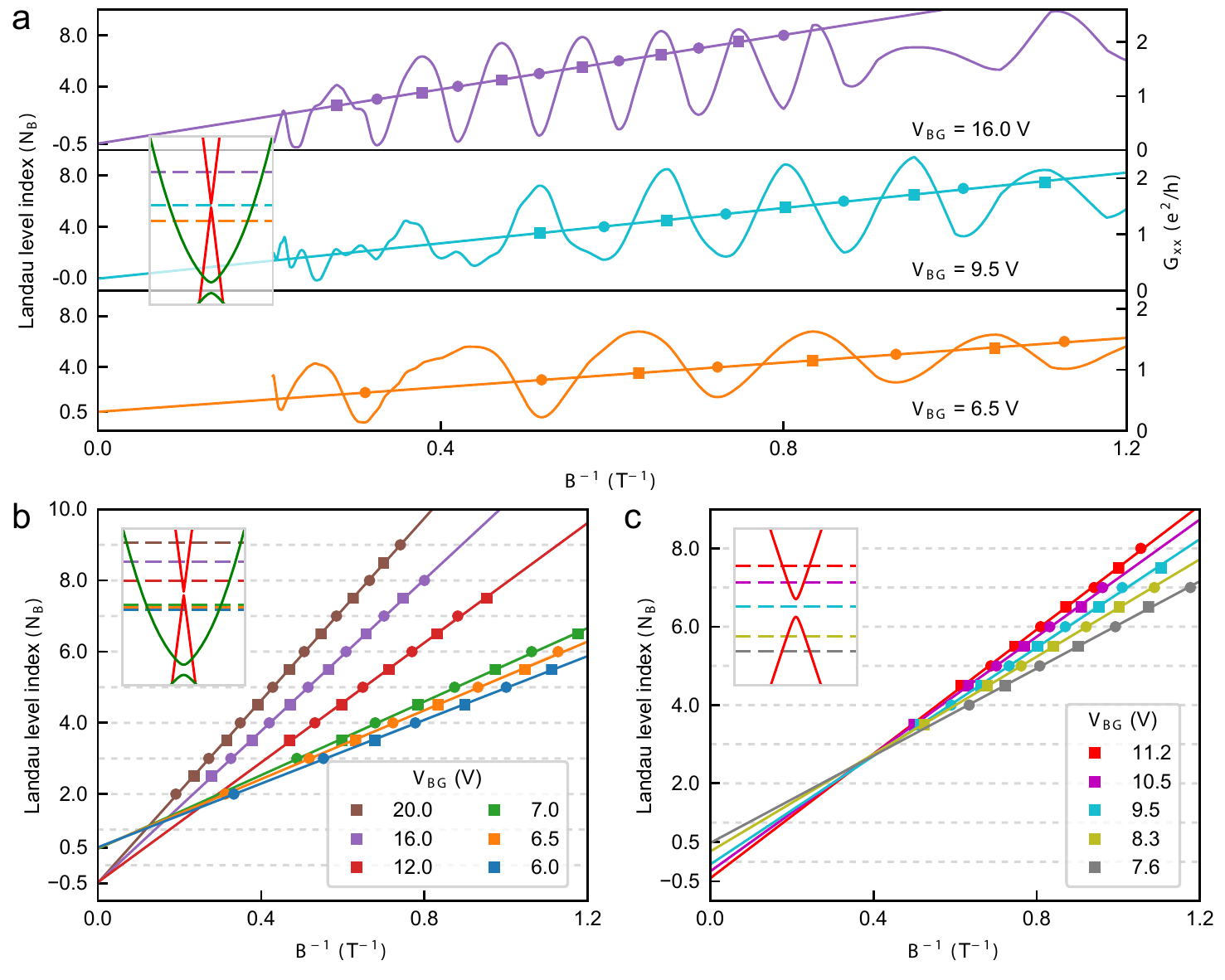}
\caption{ \label{fig:fig3}{\footnotesize  \textbf{Anomalous SdH phase shift.} (a) SdH oscillations (G$_\mathrm{xx}$) and the LL index vs inverse magnetic field fit below the band gap (green), in the band gap (red) and above the band gap (blue). Circles and squares denote the SdH minima and maxima respectively. Inset of all the panels shows the band diagram and the Fermi energy locations for which the SdH fits are shown. (b) LL index vs inverse magnetic field fits at different densities away from the band gap.  The linear fit produces $\pm\frac{1}{2}$ intercept when Fermi level lies in the MLG-like valence band and MLG-like conduction band respectively. (c) LL index vs inverse magnetic field fits at different densities close to the band gap. This shows that the intercept varies continuously from 1/2 to -1/2 when the Fermi level goes from the valence to the conduction MLG-like band by tuning the density. Inset shows the zoomed-in band diagram very close to the band gap.}}
\end{figure*}

\begin{figure*}
\includegraphics[width=16cm]{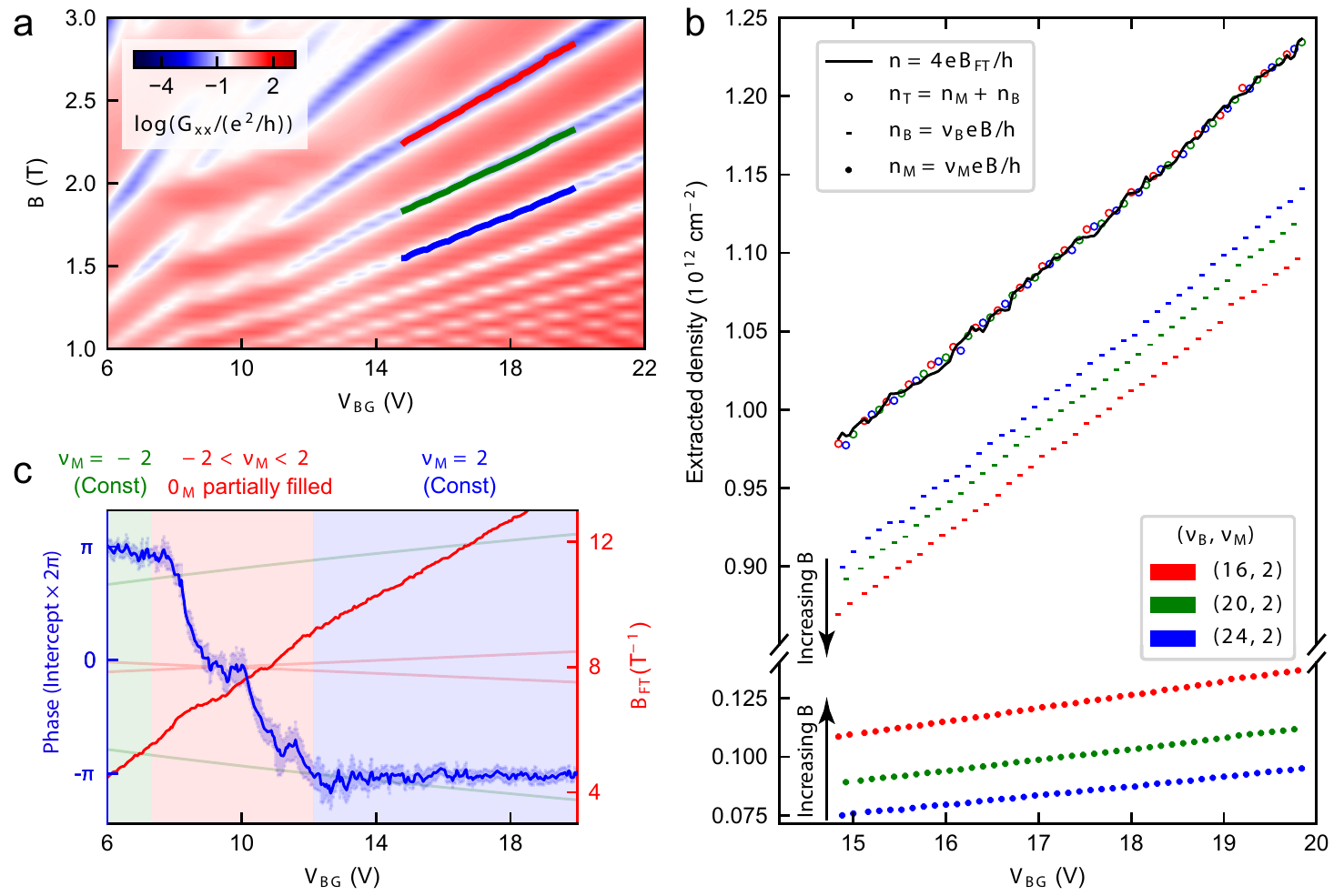}
\caption{ \label{fig:fig4} {\footnotesize  \textbf{Comparing the band specific density and the extracted density from the fits.} (a) Zoomed in measured LL fan diagram showing three lines drawn for three filling factors along which we extract the density. (b) MLG-like band density ($n_\mathrm{M}=\nu_\mathrm{M}eB/h$) and BLG-like band density ($n_\mathrm{B}=\nu_\mathrm{B}eB/h$) are marked with filled circle and dash respectively for total filling factor $\nu_\mathrm{T}$=18 (red), 22 (green) and 26 (blue). The unfilled circles of corresponding colors show the total density $n_\mathrm{T}=n_\mathrm{M}+n_\mathrm{B}$ for each total filling factors. Surprisingly, at a constant gate voltage with increasing magnetic field density of the MLG-like and the BLG-like band increases and decreases respectively, keeping total density constant at all magnetic fields (filling factors). The black line shows the density calculated from the SdH frequency ($n_\mathrm{T}$=$\frac{4e}{h}\times B_\mathrm{FT}$) which is identical to the combined density for all filling factors. (c) Intercept (blue) and slope (red) of the BLG-like SdH oscillation fit as a function of $V_\mathrm{BG}$. Fitting errors are shown for the intercept. Corresponding band diagram is overlaid for visualization. Three regions are shaded with three different colors -- green shade and blue shade indicate completely empty and completely filled $N_\mathrm{M}$=0 LL respectively whereas the saffron shade indicates partially filled $N_\mathrm{M}$=0 LL. This observed phase variation enables us to directly visualize the population/depopulation of the $N_\mathrm{M}$=0 LL.}}
\end{figure*}

%\textbf{7. More clearly extracted}

We quantify the anomalous phase shift via a detailed analysis of the SdH oscillations using the standard extrapolation method~\cite{zhang_experimental_2005}. Briefly, this involves fitting a line to the LL index (N) corresponding to a minimum in the $G_\mathrm{xx}$ vs. the corresponding inverse magnetic field ($\frac{1}{B_N}$) plot and examining the intercept at $\frac{1}{B}=0$. The method of determining LL indices is described in Supplementary Materials~\Rmnum{2}). From the intercept in the LL index axis (Fig.\ref{fig:fig3}a), we see that the intercept is 0.5 (-0.5) in the valence (conduction) band and is zero in the middle of the band gap. The 0.5 (-0.5) value of the intercept corresponds to a $\pi$ (-$\pi$) phase shift of the SdH oscillations when the Fermi level lies away from the band gap even though the phase is extracted \textit{only} from the BLG-like SdH oscillations. Fig.\ref{fig:fig3}b shows fits at several densities away from the gap (firmly in either conduction or valence band). While possessing different slopes, their intercepts assume only two quantized values: 0.5 or -0.5 depending on the Fermi energy inside the valence or conduction MLG-like Dirac cone. This reinforces the robustness of the anomalous phase shift.

Strikingly, it is only when the Fermi energy is tuned through the MLG-like band's gap, that the intercept varies continuously from 0.5 to -0.5, see Fig.\ref{fig:fig3}c. We note the smooth gate-tuning through the bandgap is possible due to the gapless nature of the BLG-like conduction bands throughout the region of interest. Both the non-trivial values and tunable nature of the anomalous phase shift sharply departs from the traditional understanding of quantum oscillation being purely sensitive to the specific Fermi surface it is sampling -- BLG-like band in the present case.

%\textbf{8. Filling-enforced phase shift}

We now focus on the origin of the anomalous phase shift. In general, SdH oscillations depend on contributions from the Fermi surfaces of both the bands: $\Delta G_{\mathrm{xx}}=G_\mathrm{M} \cos [2\pi (\frac{B_\mathrm{FM}}{B} + \gamma_\mathrm{M})]+G_\mathrm{B} \cos [2\pi (\frac{B_\mathrm{FB}}{B}+\gamma_\mathrm{B})]$, where $M$ and $B$ subscripts denote MLG-like and BLG-like bands respectively. As we explain below, the complex pattern of band fillings across multiple bands of distinct type (encoded in $(B_{\rm FB}, B_{\rm FM})$) control the SdH oscillations.

To unravel the pattern in ABA-trilayer graphene, there are two key effects to understand. First,
MLG-like LLs possess large LL separation (first LL gap is $\sim$50~meV at 2~T) even at small magnetic fields. In contrast, the LL spacing of BLG-like LLs is far smaller ($\sim$5~meV at 2~T). This means that multiple BLG-like LLs can be swept through (over large density and magnetic field windows) while keeping the filling of the MLG-like LLs constant in our experiment, see Fig.\ref{fig:fig2}d.
This is most prominent between the $N_M=0$ and $N_M=1$ MLG-like LLs, where we were able to easily resolve and analyze $\sim$10 BLG-like LLs. Even though the filling factor of the BLG-like LLs steadily varies over this region, the filling factor of the MLG-like band remains {\it pinned} to 2 due to the particularly large first MLG-like LL energy spacing and the non-field-dispersive nature of the $N_M = 0$ LL. As a result, in between MLG-like LLs [for e.g., that realized in the region E(0$_\mathrm{M})<E_\mathrm{F}<E(1_\mathrm{M}$)]  MLG-like oscillations are frozen, and the
the SdH oscillations are dominated by the BLG-like band: $\Delta G_{\mathrm{xx}}\approx G_B \cos [2\pi (\frac{B_\mathrm{FB}}{B}+\gamma_B)]$.

Second, in SdH oscillation measurements, the total density is fixed (set by the gate voltage) while the magnetic field is varied. In ABA-trilayer graphene, the total density ($n_\mathrm{T}=n_\mathrm{M}+n_\mathrm{B}$) is comprised of the individual band densities in each of the MLG-like ($n_M$) and the BLG-like ($n_B$) bands, which may reconfigure with the magnetic field while keeping $n_{\rm T}$ constant. This constraint strongly influences the BLG-like SdH oscillations. To see this we express its oscillation frequency in terms of the total density via: $B_\mathrm{FB}=\frac{n_\mathrm{B} h}{4e}=\frac{(n_\mathrm{T}-n_\mathrm{M})h}{4e}=B_\mathrm{FT}-\frac{\nu_\mathrm{M}B}{4}$, where $B_\mathrm{FT} = \frac{n_Th}{4e}$ and $\nu_M = \frac{n_Mh}{eB}$ is the filling factor of the MLG-like band. Crucially, for E(0$_\mathrm{M})<E_\mathrm{F}<E(1_\mathrm{M}$) (above the MLG-like band gap) only N$_\mathrm{M}$=0 electron like LL is filled, so the filling factor of the MLG-like band remains pinned to 2. This yields a BLG-like oscillation frequency as $\frac{B_\mathrm{FB}}{B}=\frac{B_\mathrm{FT}}{B}- 1/2$. Similarly, for E(-1$_\mathrm{M})<E_\mathrm{F}<E(0_\mathrm{M}$) (below the MLG-like band gap) the filling factor of the MLG-like band remains pinned to -2 producing $\frac{B_\mathrm{FB}}{B}=\frac{B_\mathrm{FT}}{B}+ 1/2$. Incorporating both cases into the BLG-like SdH oscillations, we obtain
 \begin{equation}\label{equn1}
 \Delta G_{\mathrm{xx}}\approx G_B \cos [2\pi (\frac{B_\mathrm{FT}}{B}+\gamma_B\pm 1/2)],
 \end{equation}
that displays an anomalous, non-trivial, and tunable phase shift, acquired due to the strong filling-enforced constraint above and below the bandgap. This yields an additional $\pi$ ($-\pi$) phase shift in the BLG-like oscillations due to the fully-emptied (fully-filled) MLG-like lowest $N_M=0$ LL. We have also extracted this phase from the theoretically calculated density of states which supports our experimental finding (see Supplementary Materials~\Rmnum{3} and ~\Rmnum{4}). In general, the filling enforced phase of the BLG-like SdH oscillations can be extracted when multiple LLs from the MLG-like bands are filled
(see Supplementary Materials~\Rmnum{5}).

%\textbf{9. Corroborating result of QO oscillation period}

The filling-enforced constraint is further corroborated by the measured quantum oscillation frequency. In particular, Eq.~(\ref{equn1}) indicates that the BLG-like quantum oscillations have $1/B$ frequency that scale with the combined density of the MLG-like and the BLG-like bands; their sum -- the total density -- is set globally by the gate voltage. We will illustrate this by focussing on the region between $N_M=0$ and $N_M=1$ MLG-like LLs. To proceed, we first note that the density in each of the bands depends on both filling factor $\nu_{B,M}$ and B. On Hall plateaus, the filling factor takes on precise quantized values, e.g., on the solid lines in Fig.~\ref{fig:fig4}a the filling factors in the bands are $\nu_B = 16,20,24$ whereas $\nu_M = 2$, obtained directly from Hall conductance measurements, see  Supplementary Materials. Strikingly, BLG-like band density $n_B$ on these lines decrease with the magnetic field (colored dash plots in Fig.~\ref{fig:fig4}b); in contrast, MLG-like band density in this region increases with the magnetic field (colored filled circles). These opposite sign variations are exactly compensated in their sum $n_T = n_M+n_B$ (see colored unfilled circles) which is fixed as a function of the magnetic field as evidenced by the collapse of the $n_T$ plots on each other --- a demonstration of the intricate reconfiguration of density between MLG-like and BLG-like bands.

Perhaps most dramatic is the precise agreement of the oscillation frequency $B_F$ directly extracted from the measured SdH oscillations (equivalent density shown as a solid black line in Fig.~\ref{fig:fig4}b) and the sum of densities from the filling factors in each band (colored unfilled circles). This concordance is expected directly from Eq.~(\ref{equn1}). Together with the anomalous phase shift, these empirically display the strong effect of the filling-enforced-constraints present in our devices.

%\textbf{10. Conclusion}

Anomalous filling-enforced phases become most pronounced when the quantum oscillations of component Fermi surfaces are resolved simultaneously over a similar range of  magnetic fields. In ABA-trilayer graphene the BLG-like band has a Fermi surface area that is about 10 times larger than that of the MLG-like band (at $V_\mathrm{BG}$=-44V), enabling oscillations of both bands to occur side-by-side (Fig.\ref{fig:fig2}c). The anomalous (non-trivial) phase shifts, that we find in BLG-like bands, amount to a ``proximity''-effect for the phase of quantum oscillations. To test this, we extracted the phase shift (of the BLG-like quantum oscillations) over a fine grid as gate-voltage is tuned through the bandgap, see Fig.\ref{fig:fig4}c. This displays the smooth evolution of phase shift from $\pi \to 0 \to -\pi$ that closely tracks the smooth evolution of Berry's phase expected for the gapped MLG-like band in inversion symmetry broken ABA-trilayer graphene. Given that in typical inversion symmetry broken systems, the change of Berry's phase is concentrated close to the band edge precisely where the number of quantum oscillations is few, ``proximity'' detection (from a co-existent band with more oscillations) can provide a surprising new facility to probe non-trivial quantum geometry. Our studies could shed light also on other topological materials like Weyl semimetals~\cite{wang_anomalous_2016} that host multiple bands.

\section*{Acknowledgements:}

We thank  Jainendra Jain, Allan MacDonald, Sreejith GJ, Umesh Waghmare, Shamashis Sengupta and Sajal Dhara for helpful discussions. Biswajit Datta is a recipient of Prime Minister's Fellowship Scheme for Doctoral Research, a public-private partnership between Science \& Engineering Research Board (SERB), Department of Science \& Technology, Government of India and Confederation of Indian Industry (CII). His host institute for research is Tata Institute of Fundamental Research, Mumbai and the partner company is Tata Steel Ltd. We acknowledge Swarnajayanti Fellowship of Department of Science and Technology (for MMD), Nanomission grant SR/NM/NS-45/2016, ONRG grant N62909-18-1-2058, and Department of Atomic Energy of Government of India for support. Preparation of hBN single crystals is supported by the Elemental Strategy Initiative conducted by the MEXT, Japan and JSPS KAKENHI Grant Number JP15K21722. J.C.W.S acknowledges the support of the Singapore National Research Foundation (NRF) under NRF fellowship award NRF-NRFF2016-05.

\section*{Author contributions:}
B.D. fabricated the device and did the measurements. B.D. and P.C.A analysed the data. B.D., L.S., M.M.D. and J. C. W. S did the calculations. K.W. and T.T. grew the hBN crystals. B.D., P.C.A, J. C. W. S and M.M.D. wrote the manuscript. M.M.D. supervised the project.

%\bibliography{trilayer_manuscript3}

%

\clearpage

%-----------------------------------------------------------------------------------------------------------------------------------------------
%                                                             Supplementary Material
%-----------------------------------------------------------------------------------------------------------------------------------------------

%%%%%%%%%% Merge with supplemental materials %%%%%%%%%%
\pagebreak
\widetext
\begin{center}
\textbf{\large Supplementary Materials: Non-trivial quantum oscillation geometric phase shift in a trivial band}
\end{center}
%%%%%%%%%% Merge with supplemental materials %%%%%%%%%%
%%%%%%%%%% Prefix a "S" to all equations, figures, tables and reset the counter %%%%%%%%%%

\setcounter{equation}{0}
\setcounter{figure}{0}
\setcounter{table}{0}
\setcounter{page}{1}

\renewcommand{\theequation}{S\arabic{equation}}
\renewcommand{\thefigure}{S\arabic{figure}}
\renewcommand{\thepage}{S\arabic{page}}
\renewcommand{\bibnumfmt}[1]{[S#1]}
\renewcommand{\citenumfont}[1]{S#1}

\section{Device fabrication}

\begin{figure}[h]
\includegraphics[width=16cm]{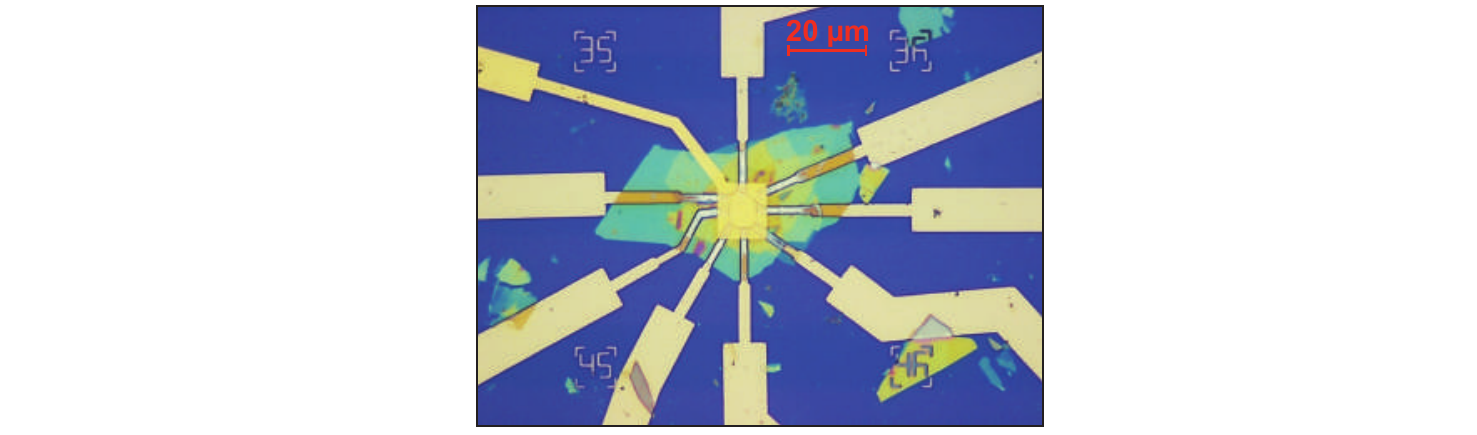}
\caption{ \label{fig:figS1}{\footnotesize Optical micrograph of our ABA-trilayer graphene device. The graphene is encapsulated by two layers of hBN. An additional hBN (for top gate insulator) was transferred on the completed device to make a uniform top gate.}}
\end{figure}

We use the Polypropylene carbonate (PPC) polymer based dry method to make the hBN-trilayer graphene-hBN stack~\cite{dean2010boron}. E-beam lithography is used to design the electrodes. Argon-Oxygen (1:1 ratio) plasma etching is used to define the  one-dimensional electrical contacts followed by the metal deposition (3~nm Chromium, 15~nm Palladium, 30~nm Gold)~\cite{wang_one-dimensional_2013}. To design a top gate we transfer one more layer of hBN as the gate insulator. The final step of e-beam lithography is done to design the metal top gate. Fig.~\ref{fig:figS1} shows the optical micrograph of the completed device.

\section{Determination of the BLG-like LL index from the experimental Hall conductance}

\begin{figure}[h]
\includegraphics[width=16cm]{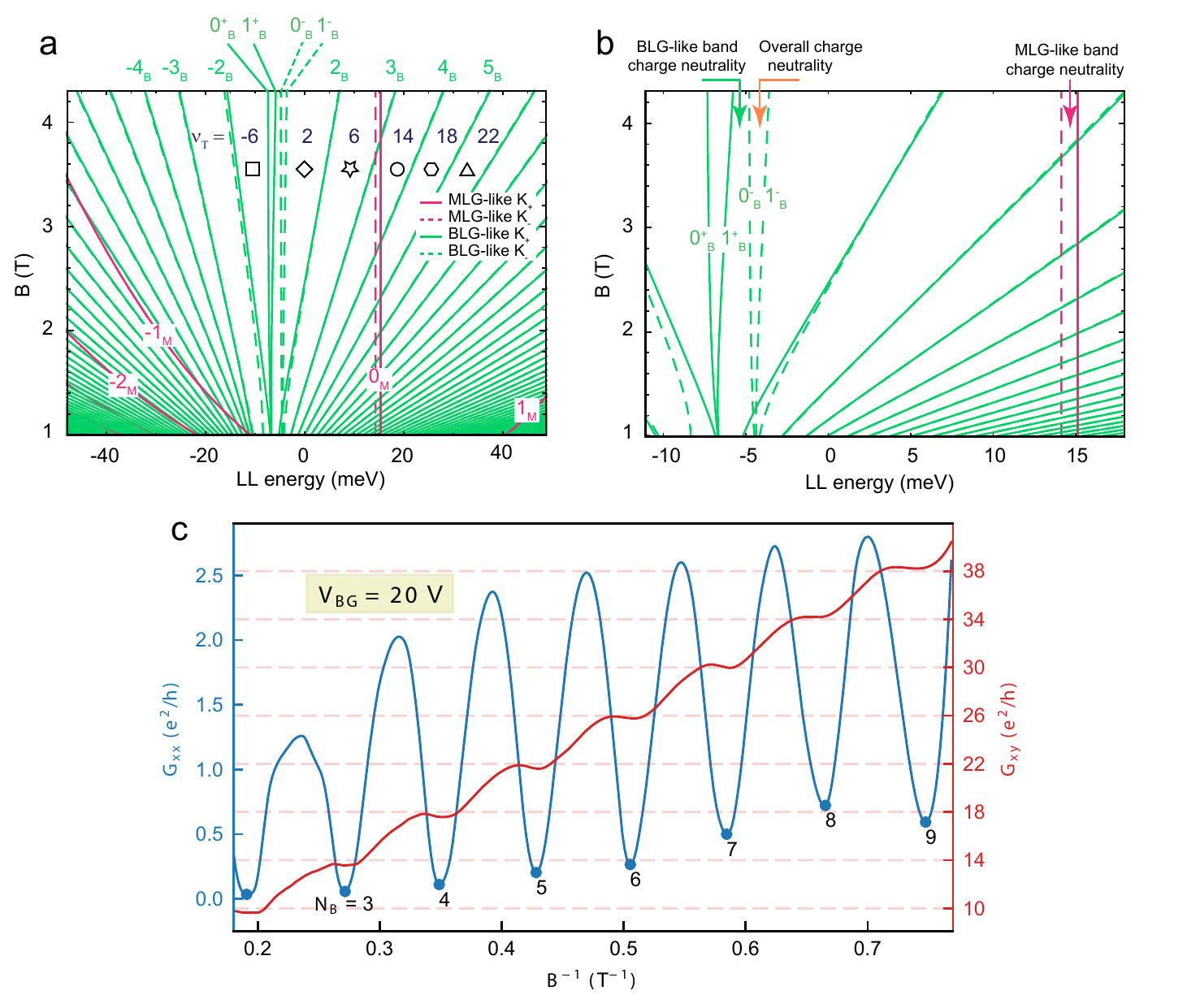}
\caption{ \label{fig:figS2} {\footnotesize \textbf{Calculation of the BLG-like LL index from the total filling factor.} (a) Different regions in the LL diagram are marked for which we show the calculation of the LL index. (b) Zoomed-in LL diagram showing the charge neutrality points of the MLG-like bands, BLG-like bands and the overall charge neutrality point of the system. Band specific filling factors are counted from their band specific charge neutrality points whereas the total filling factor is counted from the overall charge neutrality point.  (c) Experimentally measured $G_\mathrm{xx}$ and $G_\mathrm{xy}$ as a function of the magnetic field on the electron side. LL indices at all the minima are marked and are given by N$_\mathrm{B}$=$\frac{1}{4} (\nu_\mathrm{T}-2)$.}}
\end{figure}

Following previous theoretical studies~\cite{serbyn_new_2013_s}, we calculate~\cite{datta_landau_2018_s, datta_strong_2017_s} the LL energy diagram  (Fig.~\ref{fig:figS2}a) which shows that the MLG-like and BLG-like LL origins are shifted. We note that overall charge neutrality of the system is located in between 0$_\mathrm{B}$ and 1$_\mathrm{B}$ electron-like LLs (Fig.~\ref{fig:figS2}b) where Hall conductance goes to zero. Total filling factor (counted from the overall charge neutrality point) can be written as the sum of MLG-like filling factor ($\nu_\mathrm{M}$) counted from the MLG-like band origin and BLG-like filling factor ($\nu_\mathrm{B}$) counted from the BLG-like band origin: $\nu_\mathrm{T}$=$\nu_\mathrm{M}$+$\nu_\mathrm{B}$. If N$_\mathrm{M}$ is the LL index of the MLG-like LLs then filling factor of the MLG-like band above and below the band gap is given by $\nu_\mathrm{M}$=4(N$_\mathrm{M}\pm$0.5). Similarly, if N$_\mathrm{B}$ is the LL index of the BLG-like LLs then filling factor of the BLG-like band (for N$_\mathrm{B}>0$) is given by $\nu_\mathrm{B}$=4N$_\mathrm{B}$. We find the total filling factor ($\nu_\mathrm{T}$) from the experimentally measured quantized Hall conductance (G$_\mathrm{xy}$) data. Fig.~\ref{fig:figS2}c shows a line slice of the G$_\mathrm{xx}$ and G$_\mathrm{xy}$ as a function of the magnetic field at $V_\mathrm{BG}$=20~V on the electron side. Total filling factor is given by the integers where the quantum Hall G$_\mathrm{xy}$ plateaus occur. Filling factor of the MLG-like band ($\nu_\mathrm{M}$) can also be easily counted from the experimental fan diagram since the MLG-like LLs are very sparse and have a distinct parabolic dispersion. This allows us to calculate N$_\mathrm{B}$=$\frac{1}{4} (\nu_\mathrm{T}-\nu_\mathrm{M})$. Table~\ref{Tab:tabS1} shows the calculated BLG-like LL indices at different filling factors marked in Fig.~\ref{fig:figS2}a.

\begin{table}[h]
     \caption{ \label{Tab:tabS1} \textbf{Extracted LL index at 4~T for different filling factors}}
     \centering
     \begin{tabular}{*{6}{c}}\toprule

       \multicolumn{6}{c}{Below the MLG-like band gap}  \\ \midrule

      Symbol & $\nu_\mathrm{T}$ & $\nu_\mathrm{M}$=4(N$_\mathrm{M} - $0.5) & $\nu_\mathrm{B}$=4N$_\mathrm{B}$ & N$_\mathrm{M}$ & N$_\mathrm{B}$ \\  \midrule

       $\square$ &  -6 &  -2 &  -4 &  0 & -1 \\

       $\Diamond$ &  2 &  -2 &  4 &  0 & 1 \\

       $\bigstar$ &  6 &  -2 &  8 &  0 & 2 \\\midrule

        \multicolumn{6}{c}{Above the MLG-like band gap}  \\ \midrule

       Symbol & $\nu_\mathrm{T}$ & $\nu_\mathrm{M}$=4(N$_\mathrm{M} + $0.5) & $\nu_\mathrm{B}$=4N$_\mathrm{B}$ & N$_\mathrm{M}$ & N$_\mathrm{B}$ \\  \midrule

       $\fullmoon$ &  14 &  2 &  12 &  0 & 3 \\

       $\hexagon$ &  18 &  2 &  16 &  0 & 4 \\

       $\triangle$ &  22 &  2 &  20 &  0 & 5 \\  \bottomrule
     \end{tabular}
   \end{table}

%\clearpage

\section{Determination of Berry's phase from the simulated density of states (DOS) by taking a line slice at a constant energy}

\begin{figure}[h]
\includegraphics[width=16cm]{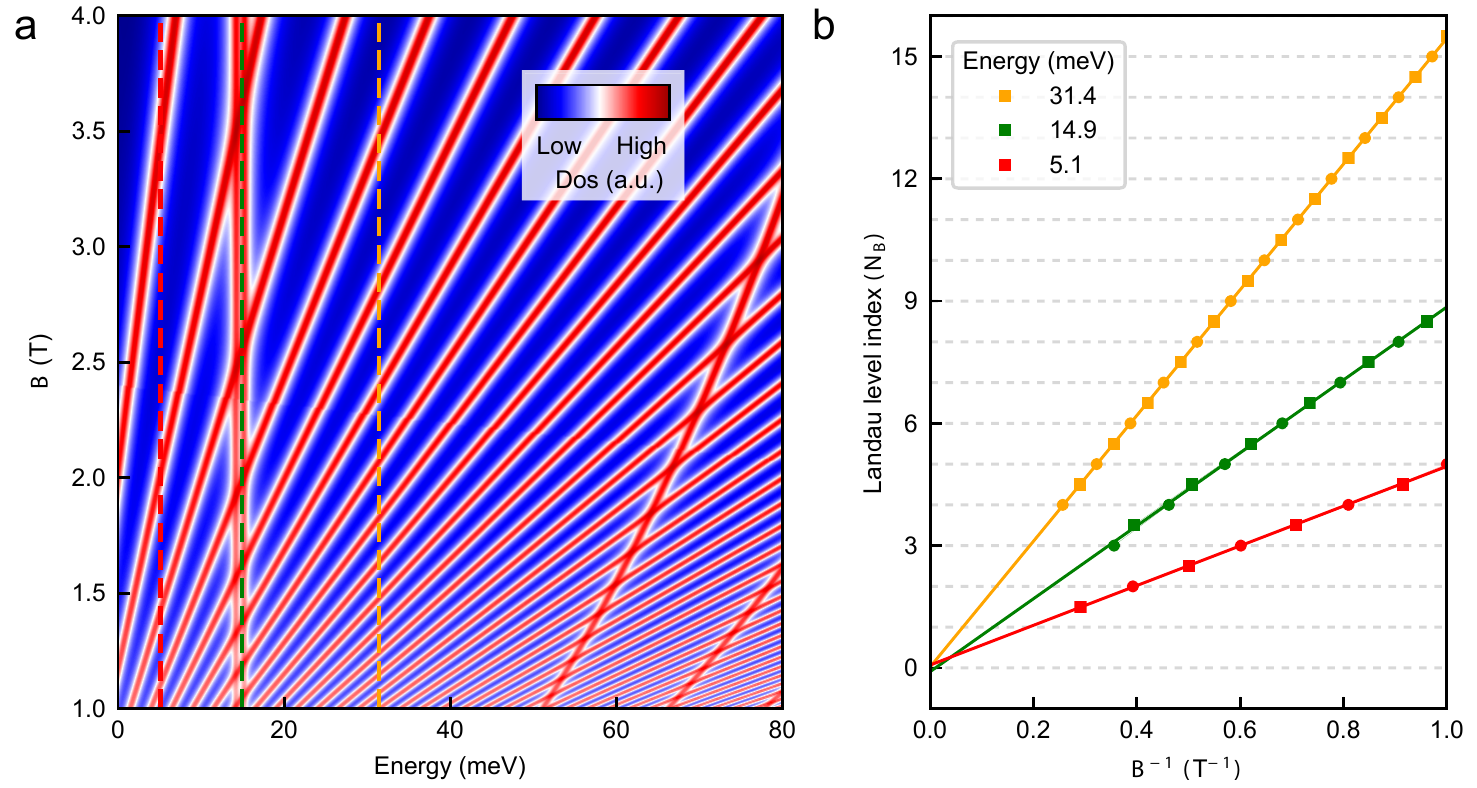}
\caption{ \label{fig:figS3} {\footnotesize\textbf{Fitting using the DOS oscillations at a constant energy.} (a) Calculated DOS as a function of energy and magnetic field. (b) BLG-like LL Fits below the band gap (red), in the band gap (green) and above the band gap (yellow).}}
\end{figure}

We extract Berry's phase also by fitting the theoretical DOS oscillations~\cite{datta_landau_2018_s, datta_strong_2017_s}. Fig.~\ref{fig:figS3}a shows the DOS as a function of Fermi energy and magnetic field. BLG-like DOS oscillation extrema are fitted at a line of constant energy. Like the experiment, integer (half-integer) LL indices are assigned at the minima (maxima) of the DOS oscillations. We note that the DOS maxima positions correspond to the experimental $G_\mathrm{xx}$ maxima. All the fits (for the Fermi level below the MLG-like band gap, in the MLG-like band gap and above the MLG-like band gap)  show zero intercepts -- irrespective of the Fermi level position in the MLG-like band (see Fig.~\ref{fig:figS3}b). This shows that the BLG-like band individually retains its $2\pi$ Berry's phase.

\section{Determination of the phase from the simulated DOS oscillation by taking a line slice at a constant density}

\begin{figure}[h]
\includegraphics[width=16cm]{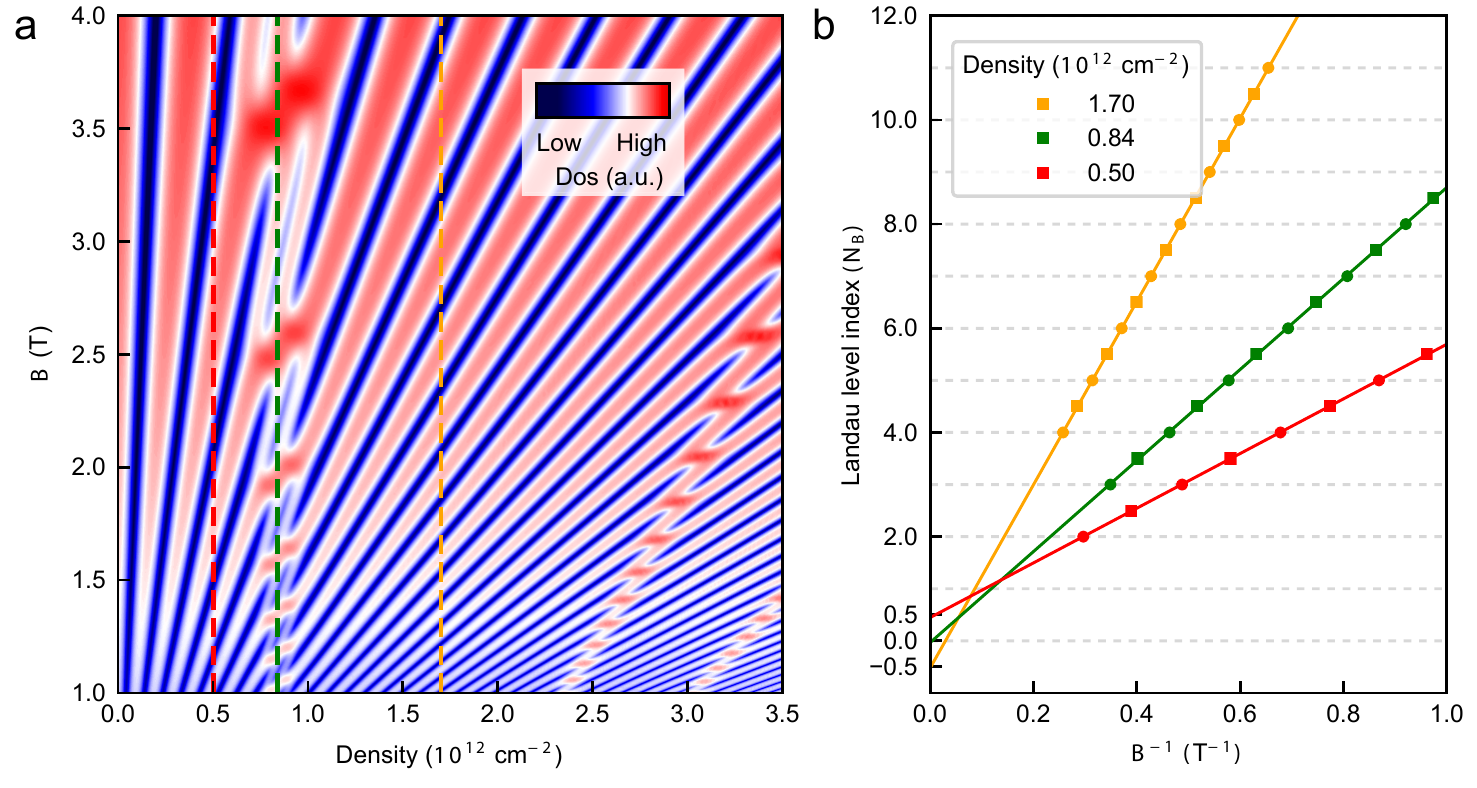}
\caption{ \label{fig:figS4}{\footnotesize \textbf{Fitting using the DOS oscillations at a constant density.} (a) Calculated DOS as a function of density and magnetic field. (b) BLG-like LL Fits below the band gap (red), in the band gap (green) and above the band gap (yellow).}}
\end{figure}

We carry out similar LL fits as shown in the main manuscript using the theoretically calculated DOS oscillations~\cite{datta_strong_2017_s,datta_landau_2018_s} (Fig.~\ref{fig:figS4}a) in the density magnetic field space. Like in the experiment, BLG-like DOS oscillation extrema are fitted at a line of constant density. We see similar anomalous phase:  $\pi$ below the MLG-like band gap and -$\pi$ above the MLG-like band gap while it goes to zero in the MLG-like band gap (Fig.~\ref{fig:figS4}b). As we have explained in the manuscript, this additional phase picked up by the trivial BLG-like Fermi surface is because of the constraint on the total density in a multiband system. This constraint naturally occurs because experimentally the SdH oscillations are measured at a constant total density controlled by the gate voltage.  The comparison between the fits done at a constant energy (Fig.~\ref{fig:figS3}b)  and at a constant density (Fig.~\ref{fig:figS4}b) clearly shows the role of density constraint to determine the phase of the quantum oscillations in a multiband system.

\section{Determination of the phase of the BLG-like SdH oscillations when multiple MLG-like LLs are filled}

\begin{figure}[h]
\includegraphics[width=16cm]{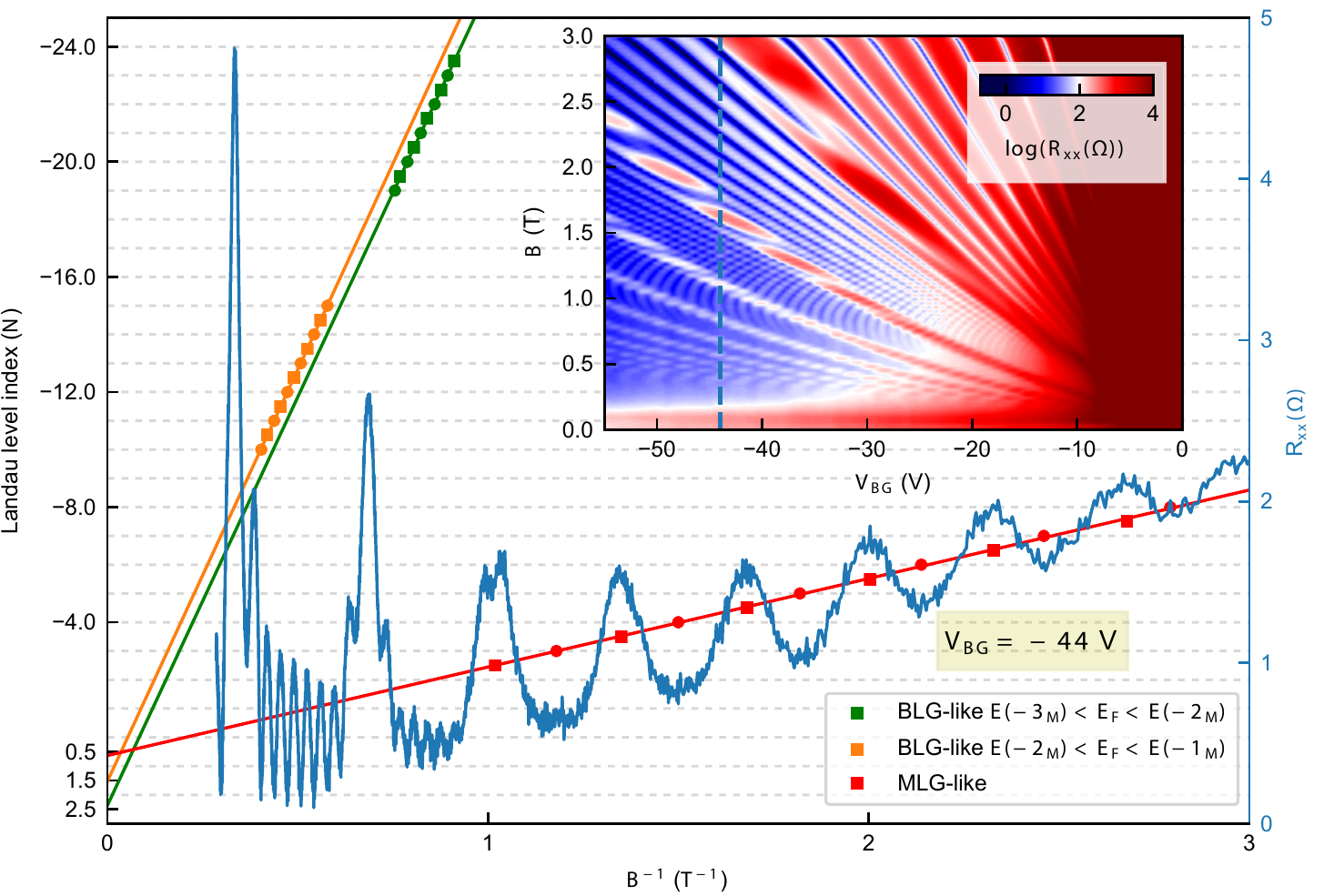}
\caption{ \label{fig:figS5} {\footnotesize Extracting the phase and SdH frequency when multiple MLG-like LLs are filled. The red line is a fit of the MLG-like LLs when the BLG-like LLs are not resolved. Orange and green lines are the fit of BLG-like LLs when -1$_\mathrm{M}$ and -2$_\mathrm{M}$ MLG-like LLs are filled respectively. Slope of the MLG-like LL fit is almost 10 times smaller than the slope of the BLG-like LL fits because the BLG-like Fermi surface area is almost 10 times larger than the MLG-like Fermi surface area for this Fermi energy.}}
\end{figure}

In general, the SdH oscillations have contributions from both the bands:
$\Delta G_{\mathrm{xx}}=G_\mathrm{M} \cos [2\pi (\frac{B_\mathrm{FM}}{B} + \gamma_\mathrm{M})]+G_\mathrm{B} \cos [2\pi (\frac{B_\mathrm{FB}}{B}+\gamma_\mathrm{B})]$. Since the first few MLG-like LLs have large gaps, it is possible that two successive MLG-like LLs contain several BLG-like LL oscillations. We fit such BLG-like LL oscillations contained between two successive MLG-like  LLs away from the crossing regions. When the Fermi energy goes through the BLG-like LL oscillations in between $N_\mathrm{M}$ and $(N+1)_\mathrm{M}$ LLs, the MLG-like filling factor remains constant to $\nu_\mathrm{M}=4(N_\mathrm{M}\pm0.5)$ because of being in the LL gap of the MLG-like LLs . Following the arguments presented in the main text for this range of Fermi energy the SdH oscillations can be captured by
 \begin{equation}
 \Delta G_{\mathrm{xx}}=G_\mathrm{B} \cos [2\pi (\frac{B_\mathrm{FT}}{B}+\gamma_B- \frac{\nu_\mathrm{M}}{4})].
 \end{equation}

In the main manuscript, we have shown the LL fits where  only the lowest MLG-like LL is filled ($\nu_\mathrm{M}=\pm2$). But, in general, the intercept of the BLG-like LL fitting depends on how many MLG-like LLs are filled.

If N and B$_\mathrm{N}$ are the LL index of the BLG-like LLs and the corresponding magnetic field at an SdH oscillation minima, then the equation of the fitting line is given by $N=\frac{B_\mathrm{FT}}{B_\mathrm{N}}+\frac{\mathrm{\Phi_\mathrm{B}}}{2\pi}$ - $\frac{\nu_\mathrm{M}}{4}$. Here the slope $B_\mathrm{FT}$ relates to the total density (n$_\mathrm{T}$=$\frac{4e}{h}\times B_\mathrm{FT}$) and the total Fermi surface area (S$_\mathrm{FT}$=$\frac{2\pi e}{\hbar}\times B_\mathrm{FT}$). Now,  only N$_\mathrm{M}$=0 and N$_\mathrm{M}$= -1 hole like LLs are filled, when the Fermi energy lies below the MLG-like band gap between the -1$_\mathrm{M}$ and -2$_\mathrm{M}$ LLs i.e.   E(-2$_\mathrm{M})<E_\mathrm{F}<E(-1_\mathrm{M}$). In this case the filling factor of the MLG-like band remains pinned to -6 making the equation of the fitting line $N=\frac{B_\mathrm{FT}}{B_\mathrm{N}}+\frac{\mathrm{\Phi_\mathrm{B}}}{2\pi}$+ $\frac{3}{2}$. Since, Berry's phase $\Phi_\mathrm{B}$=0 for BLG-like LLs, this returns 1.5 intercept at 1/B=0 (see the orange line in Fig.~\ref{fig:figS5}). Similarly, N$_\mathrm{M}$=0, N$_\mathrm{M}$= -1 and N$_\mathrm{M}$= -2 hole like LLs are filled, when the Fermi energy lies below the MLG-like band gap between the -2$_\mathrm{M}$ and -3$_\mathrm{M}$ LLs i.e.  E(-3$_\mathrm{M})<E_\mathrm{F}<E(-2_\mathrm{M}$). In this case the filling factor of the MLG-like band remains pinned to -10 making the equation of the fitting line $N=\frac{B_\mathrm{FT}}{B_\mathrm{N}}+\frac{\mathrm{\Phi_\mathrm{B}}}{2\pi}$+ $\frac{5}{2}$. This results in 2.5 intercept at 1/B=0 (see the green line in Fig.~\ref{fig:figS5}).

We also fit the MLG-like LLs at the low field when the BLG-like LLs are not resolved. At very low magnetic field $B<1$~T (i.e. 1<1/B<3 in Fig.~\ref{fig:figS5}), we resolve only MLG-like LLs since the LL spacing of the MLG-like bands are significantly larger than the BLG-like LLs. In this regime, the amplitude of the BLG-like SdH oscillations dies almost to zero, so, the SdH oscillation can be captured  only in terms of the MLG-like LLs:
\begin{equation}
 \Delta G_{\mathrm{xx}}=G_\mathrm{M} \cos [2\pi (\frac{B_\mathrm{FM}}{B}+\gamma_M)].
\end{equation}

If N and B$_\mathrm{N}$ are the LL index of the MLG-like LLs and the corresponding magnetic field, then the equation of the fitting line is given by $N=\frac{B_\mathrm{FM}}{B_\mathrm{N}}+\frac{\mathrm{\Phi_\mathrm{M}}}{2\pi}$. Here the slope $B_\mathrm{FM}$ relates to the MLG-like band density (n$_\mathrm{M}$=$\frac{4e}{h}\times B_\mathrm{FM}$) and the MLG-like Fermi surface area (S$_\mathrm{FM}$=$\frac{2\pi e}{\hbar}\times B_\mathrm{FM}$). Since, Berry's phase $\Phi_\mathrm{M}$=$\pi$ for MLG-like LLs, this returns 0.5 intercept at 1/B=0. The red line in Fig.~\ref{fig:figS5} shows that indeed the intercept of the MLG-like LLs is close to 0.5 confirming the nontrivial $\pi$ Berry's phase. This again confirms that the MLG-like band individually retains its $\pi$ Berry's phase and there is no hybridization between the bands. We note that the slope of the red line is almost an order of magnitude smaller than the orange and the green lines. This is because the Fermi surface area of the MLG-like band is roughly an order of magnitude smaller than the BLG-like Fermi surface area.

\clearpage

%\bibliography{trilayer_supp3}

%

\end{document}